# Nanoscale decoupling of electronic nematicity and structural anisotropy in FeSe thin films

Zheng Ren[1], Hong Li[1], He Zhao[1], Shrinkhala Sharma[1], Ziqiang Wang[1] and Ilija Zeljkovic[1]*

[1] Department of Physics, Boston College, 140 Commonwealth Ave, Chestnut Hill, MA 02467

* ilija.zeljkovic@bc.edu

**Abstract**

In a material prone to a nematic instability, anisotropic strain in principle provides a preferred symmetry-breaking direction for the electronic nematic state to follow. This is consistent with experimental observations, where electronic nematicity and structural anisotropy typically appear hand-in-hand. In this work, we discover that electronic nematicity can be locally decoupled from the underlying structural anisotropy in strain-engineered iron-selenide (FeSe) thin films. We use heteroepitaxial molecular beam epitaxy to grow FeSe with a nanoscale network of modulations that give rise to spatially varying strain. We map local anisotropic strain by analyzing scanning tunneling microscopy topographs, and visualize electronic nematic domains from concomitant spectroscopic maps. While the domains form so that the energy of nemato-elastic coupling is minimized, we observe distinct regions where electronic nematic ordering fails to flip direction, even though the underlying structural anisotropy is locally reversed. The findings point towards a nanometer-scale stiffness of the nematic order parameter.

**Introduction**

Electronic nematic ordering, characterized by breaking the rotational symmetry of the electronic structure, has emerged as a key signature of many unconventional superconductors [1–10]. In Fe-based superconductors, it is marked by a pronounced in-plane resistivity anisotropy [11–14], lifting of the band degeneracy [15–18] and directional scattering of electrons along a preferred Fe-Fe lattice vector [3,19–21], typically accompanied by a small orthorhombic distortion along the same direction. To gain insight into the origin of electronic nematicity, experiments have explored its evolution with chemical composition [1,11–13,22], temperature [10–19,22–24], pressure [22] and anisotropic strain [11–14,24,25]. Out of the array of these experimental handles, anisotropic strain presents a unique tuning knob that can controllably break the symmetry of the lattice. In turn, this can directly impact the overlap between inequivalent neighboring Fe-Fe orbitals, lifting the $d_{xz}$ and $d_{yz}$ orbital degeneracy, and in principle providing a preferred direction for the electronic nematicity to follow.

Microscopic imaging of Fe-based superconductors revealed the tendency of real materials to form electronic nematic domains, even in crystals under zero nominal strain [11,14,19,21,23,26–28]. The domains can have two orthogonal configurations oriented along inequivalent Fe-Fe lattice vectors. In bulk single crystals, the spatial extent of electronic nematic domains is typically at the order of a few micrometers [11,14,23,26,27]. In certain thin films however, the domain size is found to be significantly reduced compared to their bulk counterparts [21,28]. For example, while the domain size in FeSe single crystals is several micrometers [14,23], it is reduced to ~ ten nanometer length scales in thin films [21,28]. This possibly suggests that the substrate, which inevitably has a somewhat different lattice constant compared to the film, may play a role in the formation of smaller electronic nematic domains. However, quantitative measure of local symmetry-breaking strain in these systems, and its role in the development of nanoscale electronic nematic domains remains unexplored. In this work we visualize the formation of electronic nematic domains around an underlying network of structural modulations in strained multilayer films of FeSe, and discover a de-coupling of the local antisymmetric strain and electronic nematic order.

**Results**

*Observation of a network of structural modulations*

FeSe presents an excellent playground to explore the interplay of electronic nematicity and symmetry breaking strain due to its structural simplicity and the absence of magnetic ordering that is present in many other Fe-based superconductors [29,30]. In principle, various experimental methods can be used to apply strain to a material, such as voltage-controlled piezoelectric setups [11,13,25,31,32], mechanical actuators [27], differential thermal contraction [24,33] and heteroepitaxial film growth [34–37]. In this work, we use molecular beam epitaxy (MBE) to grow FeSe thin films ($a$=3.8 Å) on SrTiO$_3$(001), a substrate with a ~2% lattice mismatch ($a$=3.9 Å) (Fig. 1(a,b), Methods). We find a 2D network of modulations emerging at the surface of FeSe, propagating approximately along the Fe-Fe lattice directions (Fig. 1(c,d)). As we will subsequently show, this in turn leads to a spatially varying strain at the surface. The spatial distribution of modulation lines is qualitatively similar to those in heteroepitaxially-grown heterostructures of other chalcogenides [35–37] and arsenides [34]. In our FeSe films, ranging from 3 to 6 monolayers in thickness, this distance between neighboring modulation lines is approximately 15 to 20 nm, consistent with the spacing determined from cross-sectional transmission electron microscopy [38] and roughly consistent with the expected value based on the lattice constant mismatch between FeSe and SrTiO$_3$(001) (Supplementary Note 1) .

*Visualizing electronic nematic domains*

To characterize electronic properties of our thin films, we use low-temperature spectroscopic-imaging scanning tunneling microscopy. Although a single-unit-cell thick FeSe grown on SrTiO$_3$(001) can exhibit superconductivity with ~10-15 meV pairing gap (Supplementary Figure 6), the surface of thicker FeSe films (~a few to ~50 nm thick [21]) grown on the same substrate typically does not show superconducting behavior (Supplementary Note 8). By acquiring d$I$/d$V$(**r**,$V$) or $I$(**r**,$V$) spectra (where $I$ is the tunneling current, $V$ is the voltage applied to the sample and **r** is the relative xy-position of the tip) on a densely-spaced pixel grid, we are able to visualize spatial variations in electronic density of states as a function of energy and position. We focus on an area shown in Fig. 1(d), where we observe two striking features not immediately obvious from STM topographs (Fig. 2(b)). First, we can discern dark irregularly-shaped contours enclosing parts of the sample (denoted by white dashed lines). Second, we observe horizontal (an example denoted by green arrows) or vertical stripes (purple arrows) oriented along Fe-Fe lattice directions, with ~1.8 nm nearest-neighbor distance, which do not disperse as a function of energy in d$I$/d$V$ maps (Supplementary Figure 1). This has been interpreted as the formation of charge-stripes in the electronic nematic state [21,28], with the direction of electronic nematicity rotating by 90° across the domain boundaries. This interpretation is further supported by dispersive C$_2$-symmetric modulations pinned to individual dumbbell-shaped impurities, which also rotate by 90° across the same boundaries (Fig. 2(d)). Putting this information together, we conclude that the sample consists of two types of electronic nematic domains. We note that the smallest domains observed here are only ~100 nm$^2$, significantly smaller than those in bulk single crystals [14,23]. As we will show, the reduced domain size can be attributed to the rapidly varying strain landscape (Fig. 3).

*Strain analysis*

To measure local structural distortions, we start with an atomically-resolved STM topograph (Fig. 3(a), Supplementary Figure 3) , and apply a geometric phase analysis method [33,37] based on the Lawler-Fujita drift-correction algorithm [4]. This method allows us to determine the displacement of atoms $\mathbf{u}(\mathbf{r}) = u_a(\mathbf{r})\hat{\mathbf{a}} + u_b(\mathbf{r})\hat{\mathbf{b}}$ with picoscale resolution. The four-component strain tensor $u_{ij}(\mathbf{r}) \equiv du_i(\mathbf{r})/dr_j$ (where $i, j$ = $a, b$) can be used to extract different types of strain deformations. For example, $u_{aa}(\mathbf{r})$ represents the change of the lattice constant along the $a$-axis (relative to the average lattice constant in the field-of-view), with positive (negative) values denoting local tensile (compressive) strain. Taking into account strain along both lattice directions, it is convenient to define symmetric strain component: $S(\mathbf{r}) \equiv u_{aa}(\mathbf{r}) + u_{bb}(\mathbf{r})$ and antisymmetric strain component: $U(\mathbf{r}) \equiv u_{aa}(\mathbf{r}) - u_{bb}(\mathbf{r})$. The latter is particularly useful as a quantitative measure of structural anisotropy between the two lattice directions. We apply the strain analysis algorithm to the same area as in Fig. 2 to obtain strain maps (Fig. 3(b-e)). Tensile strain is observed along the dislocation lines in both $u_{aa}(\mathbf{r})$ and $u_{bb}(\mathbf{r})$ maps, which is sandwiched by two ribbons of compressive strain. Further away from the dislocation lines, there is tensile strain again in broader areas. To support the robustness of the strain algorithm, we note that strain maps calculated from STM topographs acquired in a range of different biases look qualitatively indistinguishable (Supplementary Figs. 2, 4). Theoretically calculated [39,40] strain maps based on a network of edge dislocations (Fig. 3(g-j)), Supplementary Note 7) also show a close resemblance to our experimental data (Fig. 3(b-e)). Therefore, we can conclude that the observed strain can be modeled well by a misfit dislocation network, with small differences that could be attributed to intrinsic orthorhombic distortion accompanying each electronic nematic domain.

*Correlation of nematic domains and antisymmetric strain*

To investigate strain inhomogeneity further, we superimpose the outlines of electronic nematic domain boundaries on top of the antisymmetric strain map *U*(**r**) (Fig. 4(c)). For the electronic nematic domain A, where the charge-stripe wave vector is oriented along the *a*-axis, it is expected that the lattice constant along the *a*-axis ($a_0$) is larger than that along the *b*-axis ($b_0$) [21]. Indeed, the average antisymmetric strain within this region is consistent with this expectation (orange color in Fig. 4(c)). Similarly, within the electronic nematic domains B, where the charge-stripe wave vector propagates along the *b*-axis, we find that on average, $b_0$ is greater than $a_0$ (purple color in Fig. 4(c)). This is consistent with the global picture revealed in elasto-resistance experiments of bulk single crystals, where electronic nematic response followed the direction of externally applied anisotropic strain [11,12]. However, our ability to probe both local anisotropy and electronic nematicity at the nanoscale enables us to explore their correlation at previously inaccessible atomic length scales.

Interestingly, we find that the distribution of antisymmetric strain values within each electronic nematic domain is highly inhomogeneous (Fig. 4(c,e)). Moreover, not only do we find variations in magnitude, but also in the sign of the anisotropic strain. In other words, within electronic nematic domain A where the direction of electronic nematicity would dictate that $a_0$ should be larger than $b_0$, we observe sizeable regions (31% of the area) where this trend is opposite (an example denoted by blue arrow in Fig. 4(c)). The same observation is also apparent in the orthogonally oriented electronic nematic domain B (yellow arrow in Fig. 4(c)), where regions with $a_0 > b_0$ comprise 21% of the whole area. Thus, our experiments reveal a local decoupling of structural and electronic anisotropy in an electronic nematic system. We emphasize that this result does not rely on the theoretical strain model in Fig. 3 or the nature of strain modulation lines. The local strain on the surface is directly determined from atomically-resolved STM topographs, and then correlated with simultaneously acquired d*I*/d*V* maps where we can see electronic nematic domains.

*Discussion*

To gain further insight into the electronic nematic domain distribution, we consider a phenomenological model where the electronic nematicity is described by an Ising order parameter field $\psi_i$. In the simplest of terms, our system can be represented as a 2D square network of lattice sites, each characterized by an electronic nematic configuration $\psi_i$ oriented along either *a* or *b*-axis. Antisymmetric strain $U(\mathbf{r}_i)$ acts as an external field that linearly couples to $\psi_i$, leading to an overall interaction energy $E_1 = -\alpha \sum_i U(\mathbf{r}_i)\psi_i$ ($\alpha > 0$), where index *i* runs over all lattice sites. If this was the only interaction in our system, to minimize the energy, the direction of $U(\mathbf{r})$ would strictly dictate the orientation of $\psi_i$ to be along the same direction. However, this is clearly contradictory to our observations (Fig. 4(f)). Therefore, we need to consider the correlation energy due to nearest neighbor interactions between the nematic fields: $E_2 = -\beta \sum_{<i,j>} \psi_i \psi_j$ ($\beta > 0$). This term accounts for the increase in the overall energy along the boundary line, i.e. the domain wall, separating two orthogonally oriented electronic nematic domains, analogous to the energy increase due to anti-alignment of nearest neighbor spins in the ferromagnetic Ising model. This model suggests that competing contributions of $E_1$ and $E_2$ will contribute to the ultimate formation of domains.

It is important to notice that the relative magnitudes of $E_1$ and $E_2$ will strongly depend on the size of electronic nematic domains formed. For micron size domains in bulk single crystals, the number of nearest neighbor pairs along a domain boundary ($N_\text{b}$) is much smaller than the total number of sites ($N_\text{total}$). Correspondingly, $E_2$ ($\propto N_\text{b}$) is negligible compared to $E_1$ ($\propto N_\text{total}$). However, as the domain size decreases, $N_\text{b}$ tends to $N_\text{total}$, and $E_2$ can become comparable to $E_1$. This can explain why no electronic domains are formed along dashed white lines in Fig. 4(f) – the energy gain from aligning $\psi_i$ with $U(\mathbf{r})$ over the small area is simply not enough to overcome the energy loss from forming a nematic boundary. Therefore, the decoupling of electronic nematicity and structural anisotropy would be energetically favored.

Our experiments highlight an application of heteroepitaxy to create a densely spaced strain grid in thin films of FeSe. We reveal a direct evidence of local decoupling between electronic nematicity and structural anisotropy, which is likely a consequence of rapidly varying anisotropic strain. Given that antisymmetric strain in our films changes sign over only ~ 5 nanometers, but the smallest electronic nematic domains observed are several times larger than that, electronic nematic rigidity length scale is likely larger than ~ 5 nanometers. This in turn suggests that it may be difficult to partition the nematic domains beyond the size already achieved here. Future experiments tracking the domain distribution as a function of temperature, could shed light on any spatial variations of $T_N$ in different strained regions and the robustness of domain boundaries with thermal cycling. Complementary to this, Te substitution for Se in strain-patterned FeSe can also allow explorations of domain formation by pushing the electronic nematic transition towards zero temperature [41]. In analogy to the magnetic field-driven motion of nematic domain boundaries detected in Ba(Fe$_{1-x}$Co$_x$)$_2$As$_2$ [42], which provided insight into substantial magneto-elastic coupling in that system, it would be interesting to investigate how the nanoscale nematic domains in FeSe behave in response to an in-plane magnetic field. Lastly, Potassium surface doping could lead to a re-emergence of superconductivity at the surface of our FeSe heterostructures [43,44], and in turn enable studying the effects of spatially varying strain on superconductivity in FeSe-based compounds.

## Methods

**MBE growth.** FeSe films were grown on Nb-doped (0.05 wt%) SrTiO$_3$ (001) (Shinkosha). The substrates were sonicated in acetone and 2-propanol, followed by annealing in O$_2$ supplied tube furnace at 1000 °C for 3 hours. This step created √13 × √13 R33.7° surface reconstruction, which has been observed in both RHEED images and STM topographs (Supplementary Figure 6). The substrates were then introduced into our MBE system (Fermion Instruments) with a base pressure of ~4x10$^{-10}$ Torr. Continuously monitored by a pyrometer, the substrates were slowly heated up to ~400 °C for growth. Fe (99%) and Se (99.999%) were co-evaporated from two Knudsen cells held at 1100 °C and 145 °C, respectively, corresponding to flux rates of 9.87x10$^{-5}$ atoms/(sec*Å$^2$) for Fe and 3.37x10$^{-3}$ atoms/(sec*Å$^2$) for Se measured by the quartz crystal microbalance. At these relatively low flux rates, it takes about 28 minutes to form each monolayer, followed by post-growth annealing at ~450 °C for 2-3 hours. After growth, the samples were either quickly transferred to the STM using a vacuum suitcase chamber held at ~1x10$^{-9}$ Torr or capped with ~50 nm thick amorphous Se layer, and de-capped in the STM chamber at ~500 °C for 2 hours. We note that FeSe films with modulations were observed in both films transferred by suitcase and de-capped thin films (Supplementary Note 4). We hypothesize that the modulations in multilayer FeSe may be related to the √13 × √13 R33.7° surface reconstruction of SrTiO$_3$ (001) [38].

**STM measurements.** STM data was acquired using a Unisoku USM1300 STM at the base temperature of ~ 4.5 K. Spectroscopic measurements were made using a standard lock-in technique with 915 Hz frequency and bias excitation as detailed in figure captions. STM tips used were home-made chemically-etched tungsten tips, annealed in UHV to bright orange color prior to STM imaging.

## Data Availability

Raw data used for the analysis shown in Figs. 2-4 can be downloaded from:
https://doi.org/10.5281/zenodo.4273119.

**Code availability**

The computer code used for data analysis is available upon request from the corresponding author.

**Acknowledgements**

The authors thank Jiun-Haw Chu for insightful discussions. I.Z. acknowledges the support from National Science Foundation grant no. NSF-DMR-1654041 (STM data and analysis) and DARPA grant no. N66001-17-1-4051 (MBE synthesis). Z.W. acknowledges support from the US Department of Energy, Basic Energy Sciences grant no. DE-FG02-99ER45747.

**Author Contributions**

MBE growth of FeSe was performed by Z.R. and S.S. STM experiments were carried out by H.L. and H.Z. Z.R., H.L., H.Z. and S.S. analyzed the STM data with the guidance from I.Z. Z.R. performed theoretical strain model calculations. Z.W. provided theoretical input on the interpretation of STM data. I.Z. supervised the project. I.Z. and Z.R. wrote the manuscript with input from all the authors.

**Competing Interests**

The authors declare no competing interests.

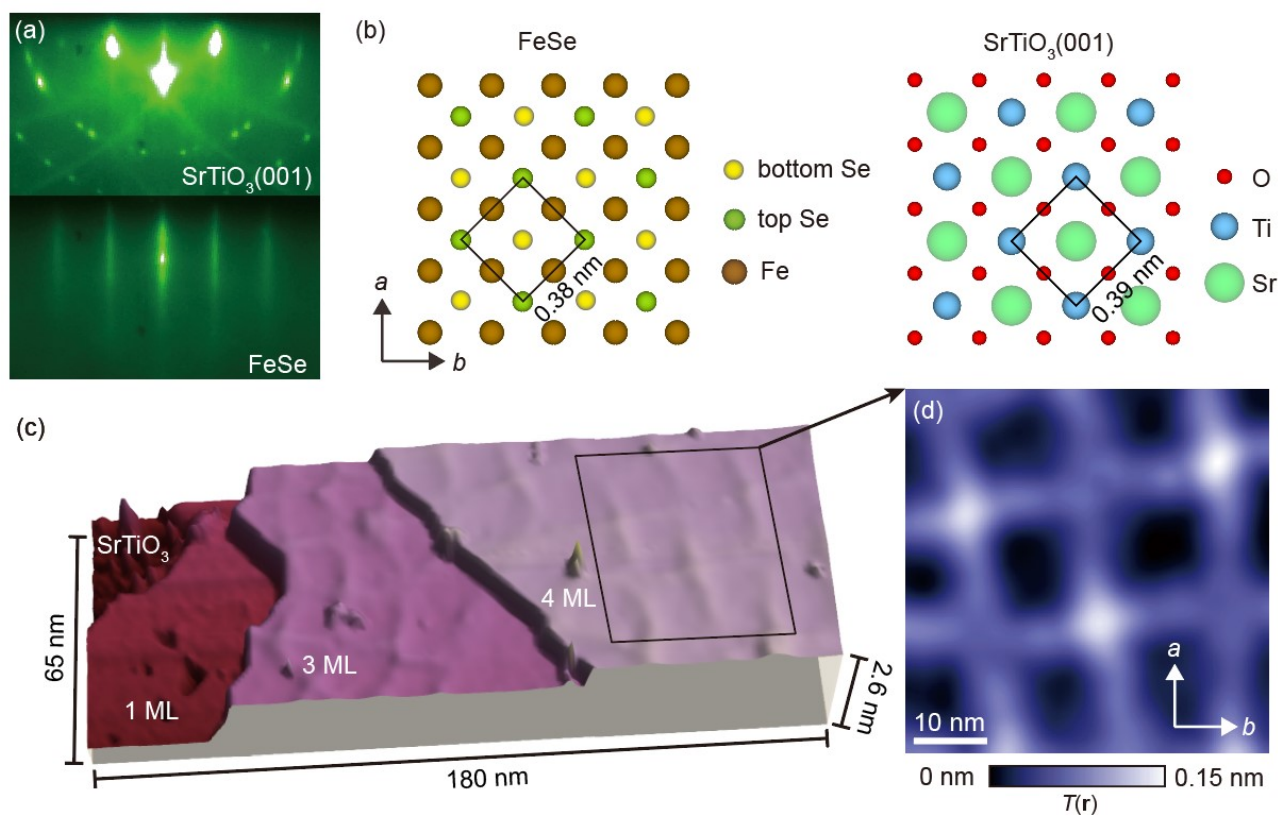

**Figure 1.** Structural characterization of FeSe thin films. (a) Reflection high-energy electron diffraction (RHEED) images (15 keV) of SrTiO$_3$(001) (top) and FeSe (bottom). (b) Top view schematic of FeSe and SrTiO$_3$ crystal structures, in which *a*- and *b*-axes denote the nearest-neighbor Fe-Fe directions. Black squares in (b) outline the Se (left) and Ti (right) unit cells. Brown, dark green and yellow spheres in the FeSe crystal structure denote Fe, top Se and bottom Se atoms, respectively. Light green, blue and red spheres in the SrTiO$_3$ crystal structure denote Sr, Ti and O atoms, respectively. (c) 3D rendered large-scale STM topograph that shows SrTiO$_3$(001) substrate, 1 monolayer (ML) FeSe, 3 ML FeSe and 4 ML FeSe terraces. The periodic structural modulations can be more easily distinguished on the 3 ML and 4 ML FeSe layers. (d) Magnification of the region outlined by the black square in (c), showing the structural modulations in the STM topograph *T*(**r**). Box smoothing over ~0.36 nm is applied to the topograph in (d). STM setup conditions: (c) $I_{set}$ = 10 pA, $V_{sample}$ = 1 V; (d) $I_{set}$ = 10 pA, $V_{sample}$ = 600 mV.

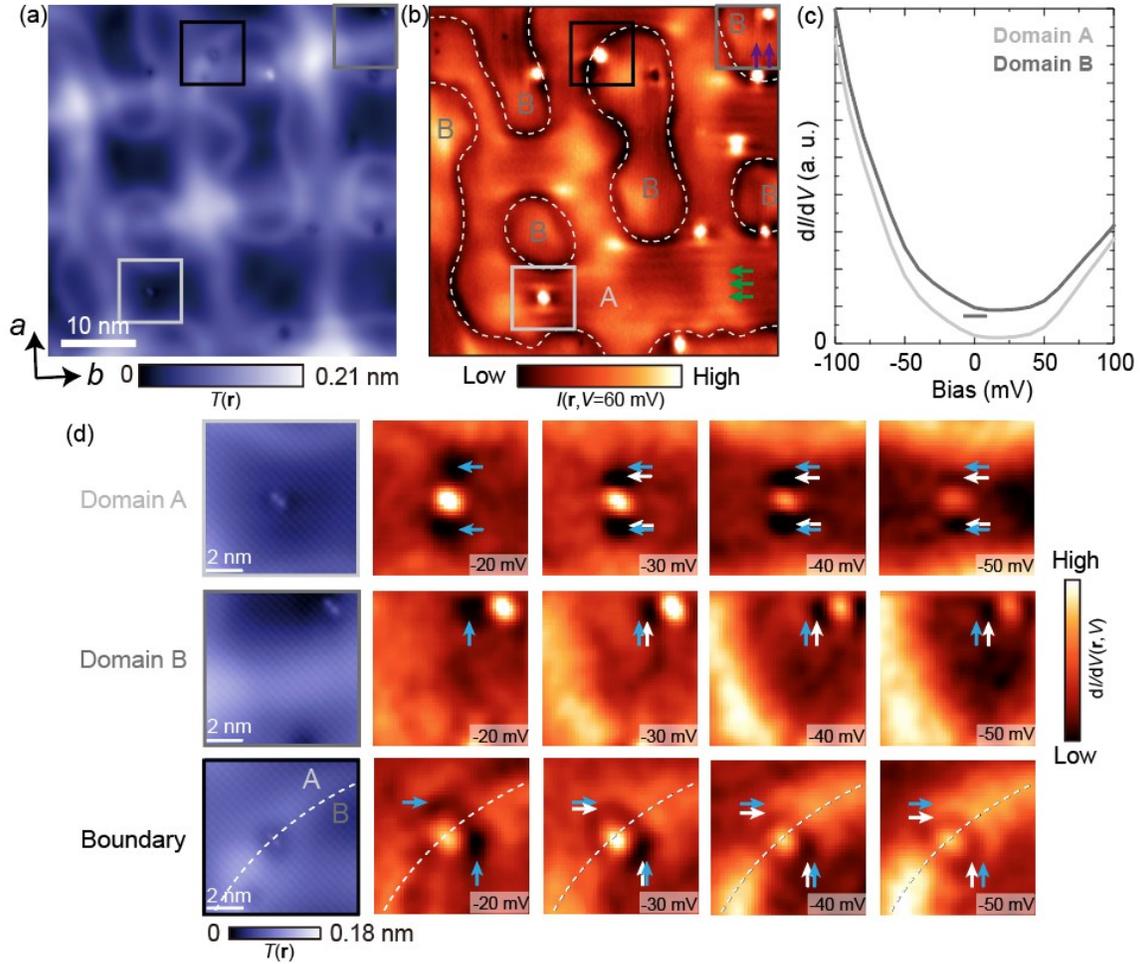

**Figure 2.** Visualizing electronic nematic domains. (a) Atomically-resolved STM topograph $T(\mathbf{r})$, and (b) tunneling current map $I(\mathbf{r}, V = 60$ mV$)$ over an identical region, showing nematic domain boundaries denoted by white dashed lines. Horizontal charge-ordered stripes (denoted by green arrows) rotate into vertical stripes (denoted by purple arrows) across the boundaries between A and B. (c) Average differential conductance d$I$/d$V$ spectra over nematic domains A and B, vertically offset for clarity. (d) Magnification of the regions near three different impurities in (a) outlined by the light gray, dark gray and black squares, which are located in the electronic nematic domain A, B and on the boundary, respectively. The last four figures in each row are d$I$/d$V(\mathbf{r},V)$ maps encompassing each impurity, showing unidirectional dispersion from -20 mV to -50 mV. The blue and white arrows serve as guides to the eye for the dispersions. We note that the direction of this signal is not dependent on the impurity shape, as all 3 impurities shown are located at the same Fe site. Symmetry of the electronic signal around impurities at the boundary of the two regions is broken down further, with one peak along each $a$- and $b$- axes, further supporting the intrinsic symmetry-broken electronic state of the system. STM setup conditions: (a) $I_{set}$ = 110 pA, $V_{sample}$ = -100 mV; (b) $I_{set}$ = 110 pA, $V_{sample}$ = -100 mV, $V_{exc}$ = 5 mV; (d) $I_{set}$ = 110 pA., $V_{sample}$ = -100 mV, $V_{exc}$ = 5 mV.

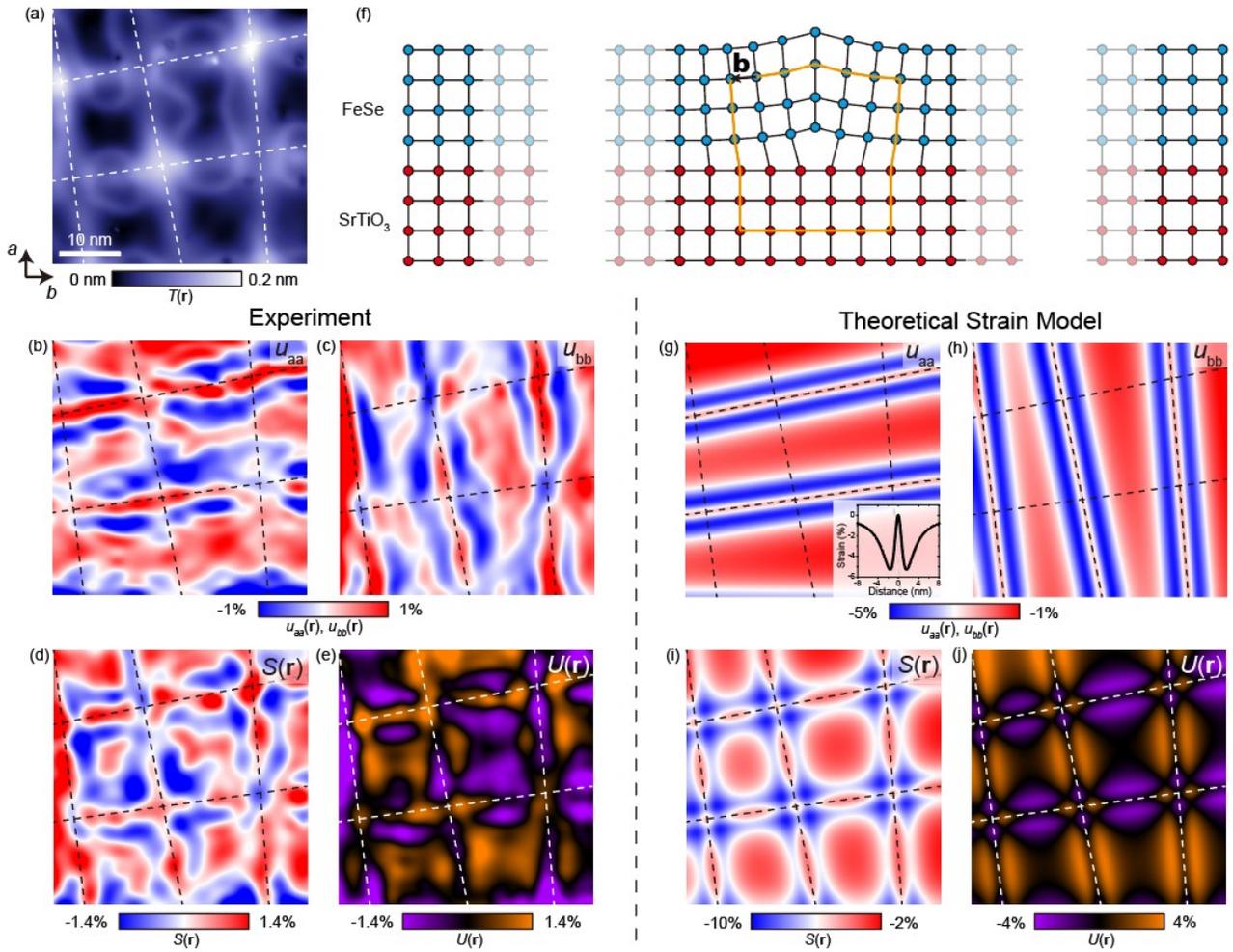

**Figure 3.** Strain analysis and comparison between experimental and theoretically predicted strain maps. (a) Atomically resolved STM topograph, where *a*- and *b*-axes denote the nearest-neighbor Fe-Fe directions. Dashed lines are guides to the eye for the structural modulations. Strain tensor components (b) $u_{aa}(\mathbf{r})$ and (c) $u_{bb}(\mathbf{r})$ derived from (a). (d) Symmetric strain map $S(\mathbf{r}) \equiv u_{aa}(\mathbf{r}) + u_{bb}(\mathbf{r})$. (e) Antisymmetric strain map $U(\mathbf{r}) \equiv u_{aa}(\mathbf{r}) - u_{bb}(\mathbf{r})$. (f) Schematic of an edge dislocation. Vector **b** denotes the Burgers vector. Blue (red) circles schematically represent the FeSe (SrTiO₃) lattice. Theoretical maps of (g) $u_{aa}(\mathbf{r})$, (h) $u_{bb}(\mathbf{r})$, (i) $S(\mathbf{r})$ and (j) $U(\mathbf{r})$. The inset in (g) shows the strain created by a single dislocation line as a function of the distance from it. In our model described in detail in Supplementary Note 7, Burgers vector **b** is set to be 0.53 nm, which is the nearest Se-Se distance on the top or the bottom of an FeSe monolayer (ML) along the dislocation direction, and the thickness *d* is set to be 1.6 nm, or 3 ML of FeSe. STM setup condition: (a) $I_{set}$ = 110 pA, $V_{sample}$ = -100 mV.

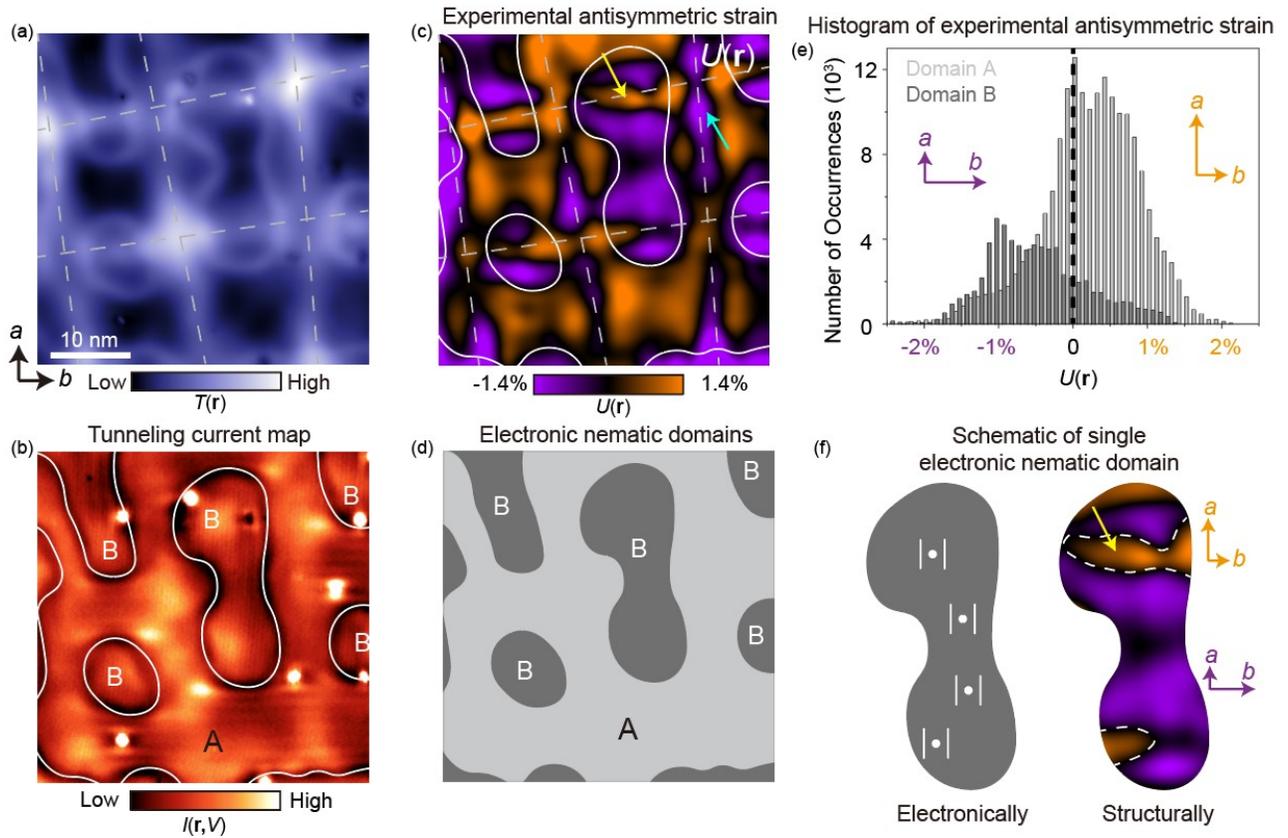

**Figure 4.** Interplay of electronic nematicity and u local strain. (a) STM topograph $T(\mathbf{r})$, and (b) tunneling current map $I(\mathbf{r}, V=60$ mV) acquired over an identical region of the sample. Dashed lines in (a,c) denote the network of structural modulations. Solid lines in (b,c) outline the electronic nematic domain boundaries. (c) The antisymmetric strain map $U(\mathbf{r})$ calculated from (a). (d) A schematic of electronic nematic domains determined from (b), where A and B refer to orthogonal electronic nematic domains. (e) Histogram of the antisymmetric strain magnitude in electronic nematic domains A and B in (c). (f) Magnification of a single nematic domain B and the antisymmetric strain distribution within it. White dots and double lines depict the Fe impurities and electronic stripes associated with the direction of electronic nematicity. Dashed lines in (f) denote the borders between regions of positive and negative antisymmetric strain. STM setup condition: (a) $I_{set}$ = 110 pA, $V_{sample}$ =-100 mV; (b) $I_{set}$ = 110 pA, $V_{sample}$ = -100 mV, $V_{exc}$ = 5 mV.

14